\documentclass[english,aps,prl,longbibliography,twocolumn]{revtex4-2}
\usepackage[T1]{fontenc}
\usepackage[latin9]{inputenc}
\usepackage{amsmath}
\usepackage{amssymb}
\usepackage{graphicx}
\usepackage{esint}

\makeatletter
\newcommand*{\Wapp} 
{3}
\newcommand*{\Wssm} 
{2} 
\newcommand*{\NPCs} 
{6} 
\newcommand*{\strfun} 
{1} 
\newcommand*{\Lorentz} 
{4} 
\newcommand*{\nalpha} 
{7} 
\newcommand*{\Fsalpha} 
{8} 
\newcommand*{\Hzero} 
{9} 

\makeatother

\usepackage{babel}
\begin{document}
\title{Exploring integrability-chaos transition with a sequence of independent
perturbations}
\author{Vladimir A. Yurovsky}
\affiliation{School of Chemistry, Tel Aviv University, 6997801 Tel Aviv, Israel}
\begin{abstract}
A gas of interacting particles is a paradigmatic example of chaotic
systems. It is shown here that even if all but one particle are fixed
in generic positions, the excited states of the moving particle are
chaotic. They are characterized by the number of principal components
(NPC) --- the number of integrable system eigenstates involved into
the non-integrable one, which increases linearly with the number of
strong scatterers. This rule is a particular case of the general effect
of an additional perturbation on the system chaotic properties. The
perturbation independence criteria supposing the system chaoticity
increase are derived here as well. The effect can be observed in experiments
with photons or cold atoms as the decay of observable fluctuation
variance, which is inversely proportional to NPC, and, therefore,
to the number of scatterers. This decay indicates that the eigenstate
thermalization is approached. The results are confirmed by numerical
calculations for a harmonic waveguide with zero-range scatterers along
its axis.
\end{abstract}
\maketitle
Evolution of integrable systems is completely predictable and, according
to the Kolmogorov-Arnold-Moser theorem (see \citep{zaslavsky}), weak
perturbations do not affect this property. There are numerous examples
of such classical and quantum systems, including stellar mechanics,
hydrogen atoms, and many-body systems, both realized in experiments,
such as quantum Newton cradle \citep{kinoshita2006,tang2018} and
cold-atom breathers \citep{dicarli2019,Luo2020} (see \citep{mazets2011,brandino2015,yurovsky2017a,Marchukov2020}
for the theoretical description) and those waiting for the realization
\citep{harshman2017}.

When the integrability-lifting perturbation is sufficiently strong,
the system evolution becomes unpredictable. Nevertheless, a completely-chaotic
system relaxes to a state described by the Gibbs ensemble, thanks
to the eigenstate thermalization mechanism --- eigenstate expectation
values are equal to microcanonical averages at the eigenstate energy
--- introduced in \citep{deutsch1991,srednicki1994} (see also \citep{rigol2008,khodja2015},
the experimental work \citep{kaufman2016}, the review \citep{deutsch2018}
and the references therein). In contrast, integrable systems relax
to states described by the generalized Gibbs ensemble\citep{rigol2007,doggen_2014,nandy2016,verstraelen2017,wu2019}.

However, a generic system lies between integrable and completely-chaotic
systems \citep{seba1990,cheon1996,legrand1997,brown2008,Yurovsky2010,Stone2010,yurovsky2011,kollar2011,Yurovsky2011b,Olshanii2012,neuenhahn2012,canovi2012,larson2013,venuti2013,fialko2014,andraschko2014,khripkov2015,venuti2015,khripkov2016,bartsch2017,dag2018,yesha2018,igloi2018,goldfriend2019,bastianello2019,huang2022,ma2022,sierant2022}.
The incomplete chaos can be related to weak integrability-lifting
perturbations \citep{kollar2011,Olshanii2012,neuenhahn2012,larson2013,huang2022,ma2022}
or phase-space separation in both classical and quantum systems. It
can have also a specific quantum nature, such as many-body localization
\citep{canovi2012,andraschko2014,sierant2022} or zero-range interactions
\citep{seba1990,cheon1996,legrand1997,Yurovsky2010,Stone2010,yurovsky2011,Yurovsky2011b}.
Chaoticity of such incompletely-chaotic systems can be characterized
by the inverse participation ratio (IPR) \citep{georgeot1997,yurovsky2011,Olshanii2012,canovi2012}.
Its inverse --- the number of principal components (NPC) --- estimates
the number of integrable system eigenstates comprising the non-integrable
one. IPR ranges from 1 for integrable systems to 0 for completely-chaotic
ones. It governs the expectation values after relaxation in incompletely-chaotic
systems with no selection rules \citep{yurovsky2011,Olshanii2012}.
This regularity has been confirmed in different systems \citep{bartsch2017}.
The fluctuations of expectation values over eigenstates are strong
for integrable systems and vanish in completely-chaotic ones, according
to the eigenstate thermalization hypothesis. These fluctuations, too,
are governed by IPR \citep{neuenhahn2012}.

Exploration of the integrability -- eigenstate thermalization crossover
is of special interest, and the means of the system chaoticity prediction
based on its Hamiltonian, with no numerical calculations, would be
very useful. The exponential decay of fluctuations in many-body systems
with the number of particles was predicted in \citep{srednicki1994}.
However, this sharp decay complicates the exploration due to the high
sensitivity of the system chaoticity to the number of particles. Analytical
predictions for single particles with random-matrix perturbations
(see \citep{Olshanii2012} and the references therein) are applicable
to strong perturbations (and chaos) only. The system chaoticity can
be modified by an additional perturbation, e.g., it can transform
an integrable system into a chaotic one. However, the effect of an
additional perturbation on a chaotic system is ambiguous, e.g., the
system integrability can be restored if the additional perturbation
is equal to the integrability-lifting one with opposite sign. The
system chaoticity increases when the additional perturbation obeys
the independence criteria determined here. Then, a linear increase
of NPC with the number of independent perturbations of the same shape
is predicted. Simultaneous decay of expectation value fluctuations
indicates that the eigenstate thermalization is approached. These
predictions are confirmed by numerical calculations for a harmonic
waveguide with zero-range scatterers along its axis. Both NPC and
expectation value fluctuations are specific characteristics of quantum
chaos and it is unclear if they have classical counterparts. 

Consider a sequence of Hamiltonians $\hat{H}_{s}=\hat{H}_{0}+\sum_{s'=1}^{s}\hat{V}_{s'}$.
Here, the integrable one $\hat{H}_{0}$ has the eigenstates $\left|\mathbf{n}\right\rangle $
and eigenenergies $E_{\mathbf{n}}$ labeled by a proper set of integrals
of motion $\mathbf{n}$. The integrability is lifted by the perturbations
$\hat{V}_{s'}$. The eigenstates $\left|\alpha_{s}\right\rangle $
of the non-integrable Hamiltonians $\hat{H}_{s}$ are labeled in the
increasing order of their eigenenergies $E_{\alpha_{s}}$. If $\hat{H}_{s}$
is invariant under some transformations, sets of $\left|\alpha_{s}\right\rangle $
with certain symmetry have to be considered separately. For example,
there may be eigenstates with certain angular momentum for rotational
symmetry of $\hat{H}_{s}$, or certain parity for inversion invariance,
or certain quasimomentum for spatial periodicity. Such sets of eigenstates
can be described by different Hamiltonians, e.g., with separated angular
momentum for rotational symmetry, or restricted to the unit cell for
spatial periodicity. Such Hamiltonians can contain fewer perturbations
than the original one $\hat{H}_{s}$. Only the case when all $\hat{H}_{s}$
have the same invariance group is considered. 

Each eigenstate $\left|\alpha_{s}\right\rangle $ can be expanded
in terms of $\left|\mathbf{n}\right\rangle $, and the strength function
\begin{equation}
W_{s}(E_{\mathbf{n}},E_{\alpha_{s}})=\left\langle \left|\left\langle \mathbf{n}|\alpha_{s}\right\rangle \right|^{2}\right\rangle \label{eq:strfun}
\end{equation}
is the probability averaged over states with a fixed energy difference
(see \citep{supplement}). The relation between the strength functions
for $\left|\alpha_{s}\right\rangle $ and $\left|\alpha_{s-1}\right\rangle $
\begin{equation}
W_{s}(E_{\mathbf{n}},E_{\alpha_{s}})\approx\sum_{E_{\alpha_{s-1}}}W_{s-1}(E_{\mathbf{n}},E_{\alpha_{s-1}})\left\langle \left|\left\langle \alpha_{s-1}|\alpha_{s}\right\rangle \right|^{2}\right\rangle \label{eq:Wssm1}
\end{equation}
is obtained (see \citep{supplement}) neglecting the interference
terms. This approximation is applicable whenever the perturbation
$\hat{V}_{s}$ is independent of $\hat{V}_{s'}$ with $s'<s$, i.e.,
\begin{equation}
\sum_{\mathbf{n},\mathbf{n}'}\left\langle \mathbf{n}|\hat{V}_{s}|\mathbf{n}'\right\rangle \left\langle \mathbf{n}'|\hat{V}_{s'}|\mathbf{n}\right\rangle \ll\sum_{s''}\sum_{\mathbf{n},\mathbf{n}'}\left|\left\langle \mathbf{n}|\hat{V}_{s''}|\mathbf{n}'\right\rangle \right|^{2}.\label{eq:Wssm1app}
\end{equation}
The summations over $\mathbf{n}$ and $\mathbf{n}'$ in the microcanonical
interval (see \citep{supplement}) lead to the Berry autocorrelation
function \citep{berry1977}. It is localized within the characteristic
de Broglie wavelength determined by the characteristic eigenstate
energy. Condition \eqref{eq:Wssm1app} is satisfied, for example,
if the spatial separation between local potentials exceeds the characteristic
de Broglie wavelength (see \citep{supplement}). (Effect of spatially-separated
perturbations on certain characteristics of energy spectra was analyzed
in \citep{langen1997}.) Other examples are the angular-dependent
potentials with no common spherical harmonics in their expansions,
such as different terms in multipole expansion, and the potentials
of different parity (see \citep{supplement}).

\begin{figure}
\includegraphics[width=3in]{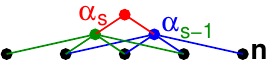}

\caption{A chart of connection between eigenstates. The number of the integrable
system eigenstates $\mathbf{n}$ connected to the non-integrable ones
increases from 4 to 5 due to an additional perturbation. \label{fig:alphassm1n}}
\end{figure}
The relation \eqref{eq:Wssm1} means that an addition of an independent
perturbation increases the number of the integrable system eigenstates
involved to the non-integrable one (see Fig. \ref{fig:alphassm1n}).
This intuitive picture illustrates the quantitative relation presented
below.

Since $W_{s}(E_{\mathbf{n}},E_{\alpha_{s}})$ should decay as $\left(E_{\mathbf{n}}-E_{\alpha_{s}}\right)^{-2}$
in the limit $\left|E_{\mathbf{n}}-E_{\alpha_{s}}\right|\rightarrow\infty$
, (see \citep{supplement}) but has no singularities as $E_{\mathbf{n}}$
and $E_{\alpha_{s}}$ never coincide, the Lorentzian profile 
\begin{equation}
W_{L}(E,\Gamma)=\frac{1}{\pi}\frac{\Gamma}{E^{2}+\Gamma^{2}}\label{eq:Lorentz}
\end{equation}
is a natural choice for the continuous strength function $W_{s}(E_{\mathbf{n}},E_{\alpha_{s}})\thickapprox W_{L}(E_{\alpha_{s}}-E_{\mathbf{n}},\Gamma_{s})\Delta E$.
Here $\Delta E$ is the average difference between eigenenergies in
the vicinity of $E_{\alpha_{s}}$. Such a strength function has been
applied to systems with strong random-matrix perturbations \citep{wigner1955,fyodorov1995,frahm1995,jacquod1995,Olshanii2012},
when the profile contains many energy levels and $\Gamma$ can be
evaluated using the Fermi golden rule. On the integrability-chaos
crossover, explored here, the profile may contain only a few levels
and the Fermi golden rule can be inapplicable, but the strength function
with some $\Gamma$ retains the necessary properties. The averaged
IPR $\eta_{s}\equiv\sum_{\mathbf{n}}\overline{\left|\left\langle \alpha_{s}|\mathbf{n}\right\rangle \right|^{4}}^{\alpha_{s}}$
(where the overbar means the microcanonical average over the states
$\alpha_{s}$) is related to the Lorentzian width $\Gamma_{s}$ as
(see \citep{supplement})
\begin{equation}
\eta_{s}=\left\{ \begin{array}{c}
2\\
3
\end{array}\right\} \frac{\Delta E}{2\pi\Gamma_{s}},\label{eq:etasGamma}
\end{equation}
where the factor $3$ is chosen for the time reversal (T) invariant
and PT invariant systems (where P is the inversion) and $2$ is chosen
otherwise. For the fixed integrable Hamiltonian $\hat{H}_{0}$ determining
the energy difference $\Delta E$ Eq. \eqref{eq:etasGamma}, together
with the relation $\Gamma_{s}=\Gamma_{s-1}+\Gamma'$ (see \citep{supplement}),
leads to the recurrence relation for NPC $\eta_{s}^{-1}$ 

\begin{equation}
\eta_{s}^{-1}=\eta_{s-1}^{-1}+\nu.\label{eq:NPCs}
\end{equation}
The parameter $\nu$ is approximately independent of $s$ if the chaotic
properties of $\left|\alpha_{s}\right\rangle $ weakly depend on $s$
and the shape of $\hat{V}_{s}$ is independent of $s$ (as for strong
scatterers with the same strength). Then \eqref{eq:NPCs} provides
a linear dependence $\eta_{s}^{-1}=\eta_{2}^{-1}+(s-2)\nu$ of NPC
on the number of scatterers. An additional independent perturbation
increases NPC even if its value is so high that the system can be
considered as a completely chaotic one. 

\begin{figure}
\includegraphics{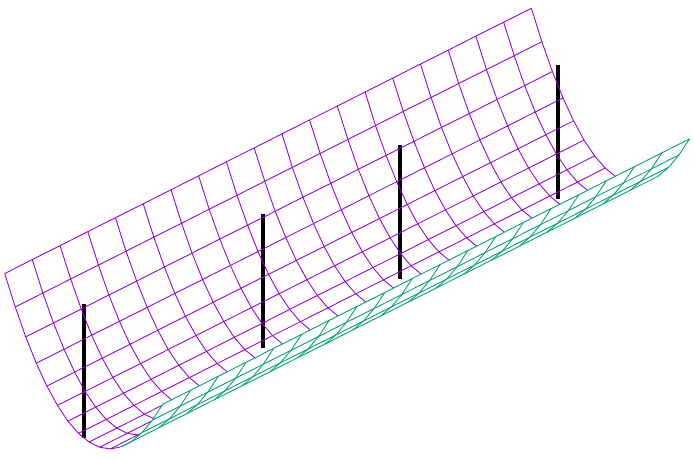}

\caption{Harmonic waveguide with four zero-range scatterers along its axis\label{fig:Harmonic-waveguide}}

\end{figure}
The prediction \eqref{eq:NPCs} is tested for models where the integrability
of a particle in a potential with separable coordinates is lifted
by $s$ fixed zero-range scatterers. Other examples of such models
are flat orthogonal billiards --- multiscatterer generalizations
of the Seba billiard \citep{seba1990}. Their energy spectrum properties
were analyzed for up to 6 scatterers \citep{cheon1996,legrand1997,yesha2018}.
Scattering in a harmonic potential was analyzed in \citep{guan2019}.
The set of scatterers is a particular case of the rank-$s$ separable
potential $\hat{V}_{s'}=V_{s'}\left|\mathcal{F}_{s'}\right\rangle \left\langle \mathcal{F}_{s'}\right|$
with the formfactors $\left|\mathcal{F}_{s'}\right\rangle $. For
such potentials the eigenstate expansion coefficients can be expressed
as (see \citep{supplement})
\begin{equation}
\left\langle \mathbf{n}|\alpha_{s}\right\rangle =\sum_{s'=1}^{s}V_{s'}\frac{\left\langle \mathbf{n}|\mathcal{F}_{s'}\right\rangle \left\langle \mathcal{F}_{s'}|\alpha_{s}\right\rangle }{E_{\alpha_{s}}-E_{\mathbf{n}}}\label{eq:nalpha}
\end{equation}
in terms of $s$ overlaps $\left\langle \mathcal{F}_{s'}|\alpha_{s}\right\rangle $
which obey the set of linear equations
\begin{equation}
\sum_{s''=1}^{s}\left(V_{s''}\sum_{\mathbf{n}}\frac{\left\langle \mathcal{F}_{s'}|\mathbf{n}\right\rangle \left\langle \mathbf{n}|\mathcal{F}_{s''}\right\rangle }{E_{\alpha_{s}}-E_{\mathbf{n}}}-\delta_{s''s'}\right)\left\langle \mathcal{F}_{s''}|\alpha_{s}\right\rangle =0.\label{eq:Fsalpha}
\end{equation}
This system has non-trivial solutions if the determinant of its matrix
is equal to zero. Then the eigenenergies $E_{\alpha_{s}}$are roots
of the determinant. High rank separable potentials can also approximate
long-range, e.g., dipole-dipole, ones \citep{derevianko2003,*derevianko2005}.
This approximation was used for energy spectra calculation \citep{kanjilal2007}. 

The present models are generalizations of the single-scatterer model
\citep{Yurovsky2010,Stone2010,yurovsky2011,Yurovsky2011b}. The integrable
Hamiltonian contains the kinetic energy and the radial harmonic potential
with the frequency $\omega_{\perp}$,
\begin{equation}
\hat{H}_{0}=\frac{\hbar^{2}}{2m}\left[\left(\frac{1}{i}\frac{\partial}{\partial z}-A\right)^{2}-\triangle_{\rho}\right]+\frac{m\omega_{\perp}^{2}\rho^{2}}{2},\label{eq:H0}
\end{equation}
where $z$ and $\rho$ are the axial and radial coordinates, respectively,
$m$ is the particle mass, and $A$ is a vector potential. The discrete
energy spectrum is provided either by the periodic boundary conditions
(PBC), $\left\langle z+L|\alpha_{s}\right\rangle =\left\langle z|\alpha_{s}\right\rangle $,
or by a hard-wall box, $\left\langle z=0|\alpha_{s}\right\rangle =\left\langle z=L|\alpha_{s}\right\rangle =0$.
The formfactors of the separable perturbation are $\left\langle \mathbf{r}|\mathcal{F}_{s'}\right\rangle =\delta_{\mathrm{reg}}(\mathbf{r}-\mathbf{R}_{s'})$,
where $\delta_{\mathrm{reg}}(\mathbf{r})$ is the Fermi-Huang pseudopotential,
and the scatterer position $\mathbf{R}_{s'}=(0,0,z_{s'})$ has the
zero radial component (see Fig. \ref{fig:Harmonic-waveguide}). The
Hamiltonian $\hat{H}$ is rotationally symmetric along the waveguide
axis and the perturbation affects only the states with zero angular
momentum. Then, only products $\left|nl\right\rangle $ of the axially
symmetric wave function $\left|n\right\rangle $ of two-dimensional
harmonic oscillator and (for PBC) a plane wave with the momentum $2\pi\hbar l/L$
are considered here. For hard-wall box, the standing waves with the
momentum $\pi\hbar l/L$ replace the plane waves (see \citep{supplement}).

Unlike a flat billiard with a constant energy density of states, in
the present model $E_{\alpha}\propto\alpha^{2/3}$, as for a three-dimensional
free particle, and the energy density of states $\partial\alpha/E_{\alpha}\propto E_{\alpha}^{1/2}$
increases with the energy. The logarithmic asymptotic freedom \citep{cheon1996}
found for flat billiards is related to decreasing effective coupling
$V_{\mathrm{eff}}\propto1/\log E$. However, it is a specific property
of the systems with $E_{\alpha}\propto\alpha$. If $E_{\alpha}\propto\alpha^{\gamma}(\gamma\neq1)$,
one can see from the derivation \citep{cheon1996} that $V_{\mathrm{eff}}\propto E^{1-1/\gamma}$
has the same energy dependence as the energy difference between the
states $\partial E_{\alpha}/\partial\alpha\propto E^{1-1/\gamma}$.
Then, the present model, as well as a generic system with $\gamma\neq1$,
does not show the asymptotic freedom.

In the absence of the vector potential, $A=0$, the energy spectrum
of the integrable Hamiltonian is degenerate, $E_{nl}=E_{n-l}$. The
degeneracy will be lifted by any potential with undefined parity.
The vector potential lifts it as well, with no complication of the
wavefunctions. However, the Hamiltonian loses the T invariance.

Four models are considered here. The non-symmetric model is T-noninvariant
and the scatterer positions $z_{1}=0$, $z_{s'}=(s'-1+\delta_{s'})L/s$
$(s'>1)$ have no symmetry due to random shifts $-0.25\leq\delta_{s'}<0.25$
chosen once for each $s$. The symmetric model with $z_{s-s'+1}=z_{s}-z_{s'}$
and equal $V_{s'}$ is PT-invariant, where P is the inversion over
$z_{s}/2$ (this model is not P-invariant, as $\hat{H}_{0}$ is not
P-invariant if $A\neq0$). The T-invariant model has $A=0$ and the
same scatterer positions as the non-symmetric one. Only this model
has degenerate $E_{nl}$. The three previous models correspond to
PBC, while the fourth, box, model, corresponds to the hard-wall box,
has $A=0$, and $z_{s'}=(s'+\delta_{s'})L/(s+1)$. 

Summation over $l$ in \eqref{eq:Fsalpha} can be done analytically
(see \citep{supplement}), leaving a sum over $\sim E_{\alpha}/(\hbar\omega_{\perp})$
values of $n$ (the closed-channel contributions with $n\hbar\omega_{\perp}>E_{\alpha}$
decay exponentially with $n$). As $E_{\alpha}\propto\alpha^{2/3}$,
calculation of the system \eqref{eq:Fsalpha} matrix and its solution
require $\sim s^{2}\alpha^{2/3}$ and $\sim s^{3}$ operations, respectively.
Then $\alpha$ eigenstates are calculated for $\alpha\gg s^{3/2}$
with $\sim s\alpha^{5/3}$ operations (cf. with $\sim\alpha^{3}$
operations required by the direct diagonalization method).

\begin{figure}
\includegraphics[width=8.6cm]{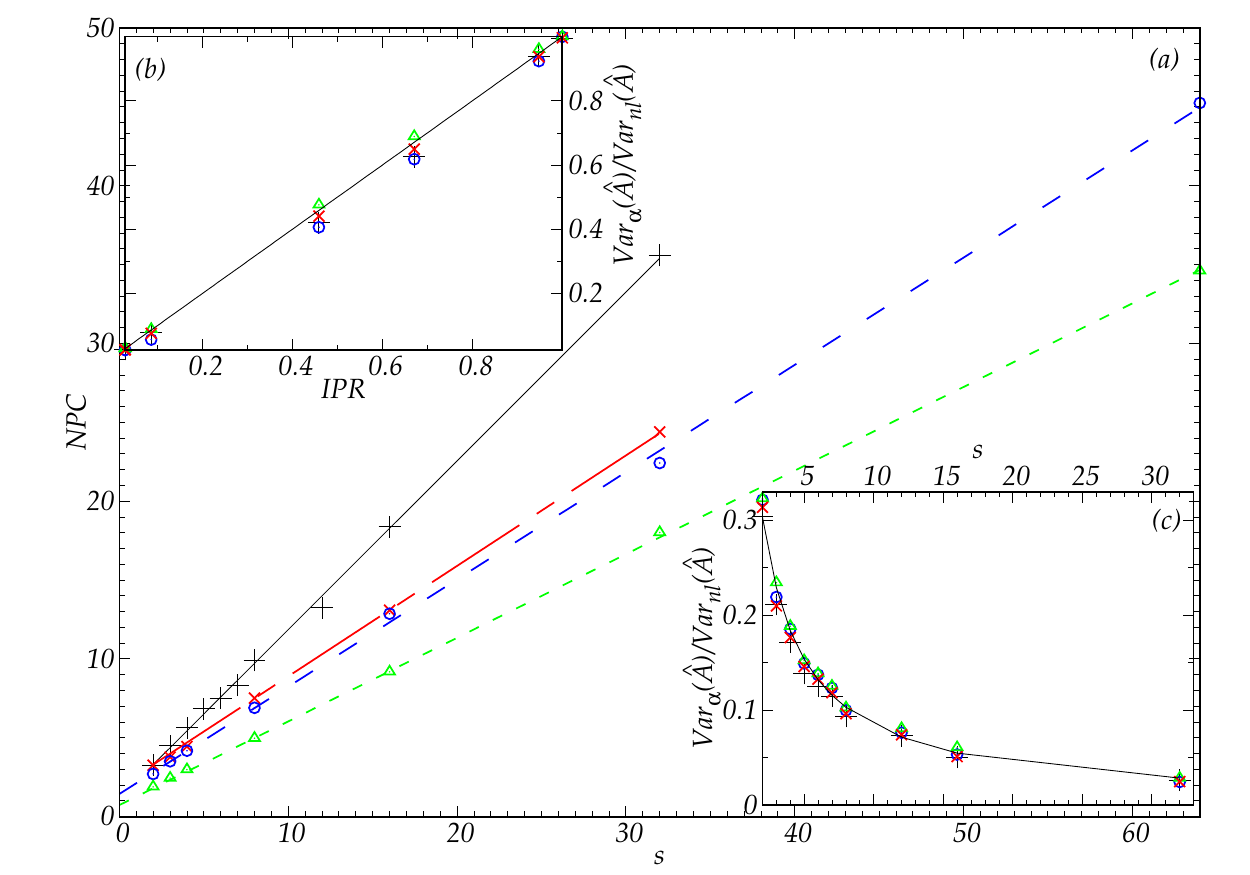}

\caption{(a) The dependence of the number of principal components on the number
of scatterers for the non-symmetric (pluses), symmetric (crosses),
box (circles), and T-invariant (triangles) models in the unitary regime
of strong perturbations. The lines represent the linear fits. (b)
The ratio of variances over eigenstates of the non-symmetric model
with 32 scatterers to ones of the integrable system as a function
of the inverse participation ratio on the change of the perturbation
strength from the unitary regime to zero. (c) The same ratio as a
function of the number of scatterers in the unitary regime. In (b)
and (c), the symbols represent four observables, namely, the axial
momentum (crosses), the part of the transverse potential energy in
the total energy (pluses), and the occupations of the positive momenta
(triangles) and of the odd axial modes (circles). The lines show IPR.\label{fig:NPCflucnscat}}
\end{figure}

The integrable system is described by two dimensionless parameters:
$\lambda=mL^{2}\omega_{\perp}/\hbar$, characterizing the aspect ratio,
and the scaled vector potential $l_{0}=LA/(2\pi)$. In the calculations,
$\lambda=\pi^{3}(1+\sqrt{5})\approx100$ and $l_{0}=0.25-e^{-4}\approx0.232$
are expressed in terms of transcendent numbers. Approximately the
same results are obtained for any $l_{0}>0.01$. In Fig. \ref{fig:NPCflucnscat}(a)
and (c), $V_{s'}=10^{6}V_{0}$ for all scatterers is in the unitary
regime ($V_{0}=2\pi\hbar^{5/2}m^{-3/2}\omega_{\perp}^{-1/2}$ is the
scale of the interaction strength). Approximately the same results
are obtained for any $V_{s'}>V_{0}$. In Fig. \ref{fig:NPCflucnscat}(b),
$V_{s'}/V_{0}=10^{6},10^{-1},10^{-2},2\times10^{-3},10^{-3},10^{-4},0$
for the $7$ IPR values from left to right (the points for $V_{s'}/V_{0}=10^{6}$
and $10^{-1}$ are almost indistinguishable). Each point in Fig. \ref{fig:NPCflucnscat}
represents an average over the states $101\leq\alpha\leq10^{6}$ of
the non-integrable system.

The plots of NPC {[}see Fig. \ref{fig:NPCflucnscat}(a){]} as a function
of the number of scatterers for the four models confirm the linear
dependence \eqref{eq:NPCs}. The linear fits have two model-dependent
parameters: $\eta_{2}$, which cannot be predicted by \eqref{eq:NPCs}
since the system with a single scatterer is not chaotic enough, and
$\nu$. In Fig. \ref{fig:NPCflucnscat}(a), $\nu=1.07,0.7,0.68$,
and $0.53$ for the non-symmetric, symmetric, box, and T-invariant
models, respectively. Taking into account the symmetry-dependent factors
in Eq. \eqref{eq:etasGamma}, we can see that $\Gamma'$ is approximately
the same for the first three models. 

The Lorentzian width is related to IPR by Eq. \eqref{eq:etasGamma}.
Figure \ref{fig:NPCflucnscat}(a) demonstrates the linear dependence
even when $\Gamma_{s}\sim\Delta E$. Therefore, even when the Fermi
golden rule is inapplicable to $\Gamma_{s}$ as the profile contains
only few energy levels, the strength function can be approximated
by the Lorentzian profile.

The physical implication of the rule \eqref{eq:NPCs} is related to
the fluctuations of expectation values $\left\langle \alpha\left|\hat{O}\right|\alpha\right\rangle $
of an observable $\hat{O}$, characterized by their variance $\mathrm{Var}_{\alpha}(\hat{O})=\overline{\left\langle \alpha\left|\hat{O}\right|\alpha\right\rangle ^{2}}-\overline{\left\langle \alpha\left|\hat{O}\right|\alpha\right\rangle }^{2}$
over the eigenstates $\left|\alpha\right\rangle $. It is proportional
to IPR and the variance for the underlying integrable system
\begin{equation}
\mathrm{Var}_{\alpha}(\hat{O})=\eta\mathrm{Var}_{\mathbf{n}}(\hat{O})\label{eq:VarAIPR}
\end{equation}
as was derived, in slightly different form, in \citep{neuenhahn2012}.
The applicability criteria of this relation can be determined using
the fact that it can also be derived in the same way as the relation
between the initial, thermal, and relaxed expectation values (7) in
\citep{yurovsky2011} replacing the density matrix by the observable.
Then \eqref{eq:VarAIPR}, like the relation (7) in \citep{yurovsky2011},
is applicable to perturbations with no selection rules. The relation
\eqref{eq:VarAIPR} was compared with numerical results \citep{neuenhahn2012}
for a many-body system with two-body interactions. However, such systems
do have selection rules, as each two-body interaction conserves quantum
numbers of other particles. It is probably the reason why the numerical
results \citep{neuenhahn2012} were described by \eqref{eq:VarAIPR}
only up to some energy-dependent factor. Systems with separable perturbations
have no selection rules, and the relation \eqref{eq:VarAIPR} describes
the dependence of variances on the number of scatterers and interaction
strength for four observables {[}see Fig. \ref{fig:NPCflucnscat}
(b) and (c){]}. The observables are: the axial momentum $\left\langle nl\left|\hat{p}_{\mathrm{ax}}\right|n'l'\right\rangle =l\delta_{n'n}\delta_{l'l}$,
the occupation of positive momenta $\left\langle nl\left|\hat{P}_{\mathrm{pos}}\right|n'l'\right\rangle =\delta_{n'n}\delta_{l'l}\theta(l)$,
the occupation of the odd axial modes $\left\langle nl\left|\hat{P}_{\mathrm{odd}}\right|n'l'\right\rangle =\delta_{n'n}\delta_{l'l}\delta_{l\mathrm{mod}2,1}$,
and the part of the transverse potential energy $m\omega_{\perp}^{2}\rho^{2}/2$
in the total energy, $\left\langle nl\left|\hat{U}\right|n'l'\right\rangle =[(2n+1)\delta_{n'n}-(n+1)\delta_{n'n+1}-n\delta_{n'n-1}]\delta_{l'l}\hbar\omega_{\perp}/(2E_{nl})$.
The averages and variances of these observable expectation values
over the integrable system eigenstates are directly calculated. The
averages are $\overline{\left\langle nl\left|\hat{p}_{\mathrm{ax}}\right|nl\right\rangle }=l_{0}$,
$\overline{\left\langle nl\left|\hat{P}_{\mathrm{pos}}\right|nl\right\rangle }=1/2$,
$\overline{\left\langle nl\left|\hat{P}_{\mathrm{odd}}\right|nl\right\rangle }=1/2$,
and $\overline{\left\langle nl\left|\hat{U}\right|nl\right\rangle }=1/3$.
Although the average expectation value of the axial momentum is constant,
its variation amplitude increases with the state energy. Then, the
variance $\mathrm{Var}_{nl}(p_{\mathrm{ax}})=mL^{2}(E_{\mathrm{max}}^{5/2}-E_{\mathrm{min}}^{5/2})/[10\pi^{2}\hbar^{2}(E_{\mathrm{max}}^{3/2}-E_{\mathrm{min}}^{3/2})]$
depends on the averaging interval $[E_{\mathrm{min}},E_{\mathrm{max}}]$
boundaries. The variances for other observables are independent of
the interval, $\mathrm{Var}_{nl}(\hat{P}_{\mathrm{pos}})=1/4$, $\mathrm{Var}_{nl}(\hat{P}_{\mathrm{odd}})=1/4$,
and $\mathrm{Var}_{nl}(\hat{U})=1/45$. The expectation values over
the non-integrable system eigenstates are calculated using the expansion
coefficients \eqref{eq:nalpha}.

Together with \eqref{eq:NPCs}, the relation \eqref{eq:VarAIPR} provides
the decay of fluctuation variance on the increase of the number of
scatterers, $\mathrm{Var}_{\alpha}(\hat{O})=\mathrm{Var}_{\mathbf{n}}(\hat{O})/[\eta_{2}^{-1}+(s-2)\nu${]},
or eigenstate thermalization approaching in a single-body system.
Then, a set of fixed scatterers mimics the behavior of many-body systems
--- sets of moving scatterers \footnote{A similar analogy between classical counterparts --- 2D problem of
fixed scatterers and the hard sphere gas --- was considered in \citep{zaslavsky}.
But to my best knowledge, no quantitative relationship between the
number of scatterers and characteristics of classic chaos is known.}. However, in many-body systems the fluctuations decay exponentially
with the number of particles. Then their chaoticity is extremely sensitive
to the number of particles and the interaction strength. For fixed
scatterers, the fluctuation decay is only inversely proportional to
their number. This opens possibilities of fine control of the system
chaoticity and exploration of the integrability-chaos crossover. 

Chaotic properties of many-body systems of interacting particles are
studied in numerous experimental and theoretical researches. However,
due to computational difficulties, a direct numerical simulation is
performed for lattice systems (e.g., \citep{rigol2007,Olshanii2012,neuenhahn2012,canovi2012,andraschko2014,khodja2015,khripkov2015,khripkov2016,nandy2016,bartsch2017,igloi2018,huang2022,ma2022})
with a finite Hilbert space, while the problem complexity increases
as a high power of the lattice site number and exponentially with
the number of particles. A single particle in an external potential
allows us to explore an infinite Hilbert space. Eigenstate thermalization
has been analyzed \citep{barnett2006} for $3\times10^{4}$ states
of a Sinai-type billiard. However, the chaoticity of that system can
not be tuned and the calculation of highly excited states is obstructed
by the increase of the coordinate grid size. The properties of systems
with several independent perturbations (particularly, scatterers)
can be tuned by the number of perturbations and their strengths. Although
the general results are confirmed by numerical calculations for a
specific model, they are applicable to any integrable system, perturbed
by several scatterers.

The predictions should be testable experimentally in several physical
systems. Tightly-trapped cold atoms of one kind can play the role
of scatterers for an atom of a second kind in a wide trap. Moreover,
several atoms of the second kind --- with interactions between them
turned off by a broad Feshbach resonance --- can be used to get averages.
In optics, photons in a cavity can be scattered by optical defects
\citep{bruck2016}.
\def\cited#1{{\hspace{\stretch{1}} \mbox{\textit{Cited:} \textbf{#1}}}}

\begin{widetext}

\renewcommand{\theequation}{S-\arabic{equation}} 
\setcounter{equation}{0} 

\section*{Supplemental Material for: \\ Exploring integrability-chaos transition
with a sequence of independent perturbations}

Numbers of equations in the Supplemental material start with S. References
to equations in the Letter do not contain S.

\section{Averages}

The microcanonical window $[E_{\mathrm{min}},E_{\mathrm{max}}]$,
containing a large number of states $\mathcal{N}_{\mathrm{MC}}$,
is small compared to the system energy scales. The microcanonical
average is defined as 
\begin{equation}
\overline{F(\alpha)}^{\alpha}=\frac{1}{\mathcal{N}_{\mathrm{MC}}}\sum_{\alpha\in\mathrm{MC}}F(\alpha),\label{eq:MCa}
\end{equation}
where $\alpha\in\mathrm{MC}$ means that $E_{\mathrm{min}}<E_{\alpha}<E_{\mathrm{max}}$.

The average over states with a fixed energy difference in the strength
function (\strfun) is defined as 
\begin{equation}
\left\langle F(\alpha,\beta)\right\rangle =\frac{1}{\mathcal{N}_{\mathrm{MC}}}\sum_{\alpha',\beta'\in\mathrm{MC}}F(\alpha',\beta')\bar{\delta}_{E_{\beta}-E_{\alpha}+E_{\alpha'},E_{\beta'}},\label{eq:FEDa}
\end{equation}
where $\bar{\delta}_{E_{\beta}-E_{\alpha}+E_{\alpha'},E_{\beta'}}$
selects the state $\beta'$ with the energy $E_{\beta'}$ closest
to $E_{\beta}-E_{\alpha}+E_{\alpha'}$.

\section{Derivation of the relation (\Wssm)}

Consider at first the Hamiltonian $\hat{H}_{2}=\hat{H}_{0}+\hat{V}_{1}+\hat{V}_{2}$.
The Schr\"odinger equations lead to the following relations between
the eigenstates $\left|\alpha_{2}\right\rangle $, $\left|\alpha_{1}\right\rangle $,
and $\left|\mathbf{n}\right\rangle $ of the Hamiltonians $\hat{H}_{2}$,
$\hat{H}_{1}$, and $\hat{H}_{0}$, respectively,
\begin{equation}
\left\langle \alpha_{1}|\alpha_{2}\right\rangle =\frac{\left\langle \alpha_{1}|\hat{V}_{2}|\alpha_{2}\right\rangle }{E_{\alpha_{2}}-E_{\alpha_{1}}},\quad\left\langle \mathbf{n}|\alpha_{1}\right\rangle =\frac{\left\langle \mathbf{n}|\hat{V}_{1}|\alpha_{1}\right\rangle }{E_{\alpha_{1}}-E_{\mathbf{n}}}\label{eq:ampot}
\end{equation}
Let us represent the expansion coefficients as $\left\langle \mathbf{n}|\alpha_{2}\right\rangle =\sum_{\alpha_{1}}\left\langle \mathbf{n}|\alpha_{1}\right\rangle \left\langle \alpha_{1}|\alpha_{2}\right\rangle $
and separate the diagonal part of the strength function (\strfun)
\begin{align}
W_{2}^{\mathrm{diag}}(E_{\mathbf{n}},E_{\alpha_{2}}) & =\left\langle \sum_{\alpha_{1}}\left|\left\langle \mathbf{n}|\alpha_{1}\right\rangle \right|^{2}\left|\left\langle \alpha_{1}|\alpha_{2}\right\rangle \right|^{2}\right\rangle \nonumber \\
 & =\frac{1}{\mathcal{N}_{\mathrm{MC}}}\sum_{E_{\alpha_{1}}}\sum_{\mathbf{n}',\alpha'_{1}\in\mathrm{MC}}\left|\left\langle \mathbf{n}'|\alpha'_{1}\right\rangle \right|^{2}\bar{\delta}_{E_{\alpha_{1}}-E_{\mathbf{n}}+E_{\mathbf{n}'},E_{\alpha'_{1}}}\frac{1}{(E_{\alpha_{2}}-E_{\alpha_{1}})^{2}}\sum_{\alpha'_{2}}\left|\left\langle \alpha'_{1}|\hat{V}_{2}|\alpha'_{2}\right\rangle \right|^{2}\bar{\delta}_{E_{\alpha_{2}}-E_{\alpha_{1}}+E_{\alpha'_{1}},E_{\alpha'_{2}}}.\label{eq:W2diag}
\end{align}
The eigenstates $\left|\alpha'_{1}\right\rangle $ are independent
of the potential $\hat{V}_{2}$ as the Hamiltonian $\hat{H}_{1}$
does not include $\hat{V}_{2}$. Besides, $E_{\alpha'_{1}}-E_{\alpha'_{2}}$
is independent of $\alpha'_{1}$ since $E_{\alpha'_{1}}$ and $E_{\alpha'_{2}}$
are independent being eigenenergies of different non-integrable Hamiltonians.
This means that the value of the last sum in Eq. \eqref{eq:W2diag}
appears as the states $\left|\alpha'_{1}\right\rangle $ were indiscriminately
chosen from a microcanonical interval. Then the sum can be approximated
by its microcanonical average over $\alpha'_{1}$, leading to
\begin{equation}
W_{2}^{\mathrm{diag}}(E_{\mathbf{n}},E_{\alpha_{2}})\approx\sum_{E_{\alpha_{1}}}W_{1}(E_{\mathbf{n}},E_{\alpha_{1}})\left\langle \left|\left\langle \alpha_{1}|\alpha_{2}\right\rangle \right|^{2}\right\rangle .\label{eq:Wdiag2}
\end{equation}
If the non-diagonal part of the strength function can be neglected,
replacing $\alpha_{2}$, $\alpha_{1}$, $\hat{V}_{2}$, and $\hat{V}_{1}$
by $\alpha_{s}$, $\alpha_{s-1}$, $\hat{V}_{s}$, and $\sum_{s'=1}^{s-1}\hat{V}_{s'}$,
respectively, we get the relation (\Wssm).

In order to evaluate the non-diagonal part, let us represent the expansion
coefficients as 
\[
\left\langle \mathbf{n}|\alpha_{2}\right\rangle =\frac{1}{E_{\alpha_{2}}-E_{\mathbf{n}}}\sum_{\alpha_{1}}\left(\left\langle \mathbf{n}|\hat{V}_{1}|\alpha_{1}\right\rangle \left\langle \alpha_{1}|\alpha_{2}\right\rangle +\left\langle \mathbf{n}|\alpha_{1}\right\rangle \left\langle \alpha_{1}|\hat{V}_{2}|\alpha_{2}\right\rangle \right)
\]
using Eqs. \eqref{eq:ampot} and the completeness of the set $\left|\alpha_{1}\right\rangle $
{[}$\alpha_{1}$ here is the same as in \eqref{eq:W2diag}{]}. The
corresponding probabilities can be expressed 
\begin{align}
\left|\left\langle \mathbf{n}|\alpha_{2}\right\rangle \right|^{2}=\frac{1}{(E_{\alpha_{2}}-E_{\mathbf{n}})^{2}}\sum_{\alpha_{1},\alpha'_{1}} & \left(B_{\alpha_{1}\alpha'_{1}}\left(\hat{V}_{1},\hat{V}_{1},\mathbf{n}\right)B_{\alpha'_{1}\alpha{}_{1}}\left(\hat{I},\hat{I},\alpha_{2}\right)+B_{\alpha_{1}\alpha'_{1}}\left(\hat{I},\hat{I},\mathbf{n}\right)B_{\alpha'_{1}\alpha{}_{1}}\left(\hat{V}_{2},\hat{V}_{2},\alpha_{2}\right)\right.\nonumber \\
 & \left.+B_{\alpha_{1}\alpha'_{1}}\left(\hat{I},\hat{V}_{1},\mathbf{n}\right)B_{\alpha'_{1}\alpha{}_{1}}\left(\hat{I},\hat{V}_{2},\alpha_{2}\right)+B_{\alpha_{1}\alpha'_{1}}\left(\hat{V}_{1},\hat{I},\mathbf{n}\right)B_{\alpha'_{1}\alpha{}_{1}}\left(\hat{V}_{2},\hat{I},\alpha_{2}\right)\right)\label{eq:nalpha2B}
\end{align}
in terms of the products of the matrix elements
\[
B_{\alpha_{1}\alpha'_{1}}\left(\hat{O}_{1},\hat{O}_{2},\beta\right)=\left\langle \alpha_{1}|\hat{O}_{1}|\beta\right\rangle \left\langle \beta|\hat{O}_{2}|\alpha'_{1}\right\rangle .
\]
 In the generic case, different degrees of freedom of $\hat{H}_{0}$
have incommensurate frequencies, and the quantum numbers $\mathbf{n}$
and $\mathbf{n}'$ of the energy-neighboring states will be mutually
uncorrelated. Then, substituting \eqref{eq:nalpha2B} into the strength
function (\strfun), we can replace $B_{\alpha_{1}\alpha'_{1}}\left(\hat{O}_{1},\hat{O}_{2},\mathbf{n}\right)$
by its microcanonical average $\overline{B_{\alpha_{1}\alpha'_{1}}\left(\hat{O}_{1},\hat{O}_{2},\mathbf{n}\right)}^{\mathbf{n}}$
{[}see \eqref{eq:MCa}{]} and obtain the following approximation for
the fixed energy difference average {[}see \eqref{eq:FEDa}{]}
\begin{equation}
\left\langle \frac{1}{(E_{\alpha_{2}}-E_{\mathbf{n}})^{2}}B_{\alpha_{1}\alpha'_{1}}\left(\hat{O}_{1},\hat{O}_{2},\mathbf{n}\right)B_{\alpha'_{1}\alpha{}_{1}}\left(\hat{O}_{3},\hat{O}_{4},\alpha_{2}\right)\right\rangle \approx\frac{1}{(E_{\alpha_{2}}-E_{\mathbf{n}})^{2}}\overline{B_{\alpha_{1}\alpha'_{1}}\left(\hat{O}_{1},\hat{O}_{2},\mathbf{n}\right)}^{\mathbf{n}}\overline{B_{\alpha'_{1}\alpha{}_{1}}\left(\hat{O}_{3},\hat{O}_{4},\alpha_{2}\right)}^{\alpha_{2}}\label{eq:FEDaBB}
\end{equation}

The independence of the random functions $\left\langle \mathbf{n}|\alpha_{s}\right\rangle $
and $\left\langle \mathbf{n}|\alpha'_{s}\right\rangle $ for $\alpha_{s}\neq\alpha'_{s}$
\citep{berry1977,srednicki1994} leads to the approximate equality
of the microcanonical averages 
\begin{equation}
\overline{|\alpha_{s}\left\rangle \right\langle \alpha_{s}|}^{\alpha_{s}}\approx\overline{|\mathbf{n}\left\rangle \right\langle \mathbf{n}|}^{\mathbf{n}}\label{eq:MCaveq}
\end{equation}
 of the product of two eigenfunctions. This results in
\begin{equation}
\overline{B_{\alpha_{1}\alpha'_{1}}\left(\hat{I},\hat{I},\beta\right)}^{\beta}\approx\frac{1}{\mathcal{N}_{\mathrm{MC}}}\delta_{\alpha_{1}\alpha'_{1}},\quad\overline{B_{\alpha_{1}\alpha'_{1}}\left(\hat{I},\hat{V},\beta\right)}^{\beta}\approx\frac{1}{\mathcal{N}_{\mathrm{MC}}}\left\langle \alpha_{1}|\hat{V}|\alpha'_{1}\right\rangle \label{eq:Balphaaver}
\end{equation}
if $\alpha_{1}$ and $\alpha'_{1}$ belong to the microcanonical interval
(otherwise, the averages vanish). Then, using Eqs. \eqref{eq:nalpha2B},
\eqref{eq:FEDaBB}, \eqref{eq:MCaveq}, and \eqref{eq:Balphaaver}
one can approximate the diagonal and non-diagonal parts of the strength
function (\strfun) as
\begin{multline}
W_{2}^{\mathrm{diag}}(E_{\mathbf{n}},E_{\alpha_{2}})\approx\frac{1}{\mathcal{N}_{\mathrm{MC}}^{2}(E_{\alpha_{2}}-E_{\mathbf{n}})^{2}}\left[\sum_{\mathbf{n},\mathbf{n}'\in\mathrm{MC}}\left(\left|\left\langle \mathbf{n}|\hat{V}_{1}|\mathbf{n}'\right\rangle \right|^{2}+\left|\left\langle \mathbf{n}|\hat{V}_{2}|\mathbf{n}'\right\rangle \right|^{2}\right)+2\sum_{\alpha_{1}\in\mathrm{MC}}\left\langle \alpha_{1}|\hat{V}_{1}|\alpha_{1}\right\rangle \left\langle \alpha_{1}|\hat{V}_{2}|\alpha_{1}\right\rangle \right]\label{eq:Wdiag2V}
\end{multline}
\[
W_{2}^{\mathrm{nond}}(E_{\mathbf{n}},E_{\alpha_{2}})\approx\frac{2}{\mathcal{N}_{\mathrm{MC}}^{2}(E_{\alpha_{2}}-E_{\mathbf{n}})^{2}}\left(\sum_{\mathbf{n},\mathbf{n}'\in\mathrm{MC}}\left\langle \mathbf{n}|\hat{V}_{1}|\mathbf{n}'\right\rangle \left\langle \mathbf{n}'|\hat{V}_{2}|\mathbf{n}\right\rangle -\sum_{\alpha_{1}\in\mathrm{MC}}\left\langle \alpha_{1}|\hat{V}_{1}|\alpha_{1}\right\rangle \left\langle \alpha_{1}|\hat{V}_{2}|\alpha_{1}\right\rangle \right).
\]
The last, single, sums here can be neglected compared to the double
sums. Then, the non-diagonal part is negligible when the condition
(\Wapp) for $s=2$ is satisfied. The recurrent application of this
condition provides the condition (\Wapp) for any $s$.

The microcanonical average of the product of two eigenfunctions 
\[
\overline{\left\langle \mathbf{r}_{1}|\mathbf{n}\right\rangle \left\langle \mathbf{n}|\mathbf{r}_{2}\right\rangle }^{\mathbf{n}}\approx C_{\mathrm{Berry}}(|\mathbf{r}_{1}-\mathbf{r}_{2}|,(\mathbf{r}_{1}+\mathbf{r}_{2})/2)
\]
is the Berry autocorrelation function \citep{berry1977}
\[
C_{\mathrm{Berry}}(r,\mathbf{R})=\varGamma(D/2)\frac{J_{D/2-1}(K(\mathbf{R})r)}{(K(\mathbf{R})r/2)^{D/2-1}},
\]
where $D$ is the number of degrees of freedom (the dimension of the
vectors $\mathbf{n}$ , $\mathbf{r}$, and $\mathbf{R}$), $\varGamma$
is the $\Gamma$-function, $J$ is the Bessel function (see \citep{DLMF}),
and $K(\mathbf{R})$ is the absolute value of the local wavevector.
The autocorrelation function decays when $r$ exceeds the characteristic
de Broglie wavelength $K^{-1},$ determined by the mean energy of
the microcanonical interval.

Consider several important cases when $W_{2}^{\mathrm{nond}}$ becomes
negligible and, therefore, the condition (\Wapp) is satisfied. For
local coordinate-dependent potentials $\hat{V}_{1}$ and $\hat{V}_{2}$
the sum in $W_{2}^{\mathrm{nond}}$ can be transformed as
\begin{equation}
\sum_{\mathbf{n},\mathbf{n}'\in\mathrm{MC}}\left\langle \mathbf{n}|\hat{V}_{1}|\mathbf{n}'\right\rangle \left\langle \mathbf{n}'|\hat{V}_{2}|\mathbf{n}\right\rangle \approx\int d^{D}Rd^{D}rV_{1}(\mathbf{R}+\mathbf{r}/2)V_{2}(\mathbf{R}-\mathbf{r}/2)C_{\mathrm{Berry}}^{2}(r,\mathbf{R}).\label{eq:sumnnond}
\end{equation}
 Then, $W_{2}^{\mathrm{nond}}$ becomes negligible when the distance
between $V_{1}$ and $V_{2}$ exceeds the characteristic de Broglie
wavelength. 

If the ranges of $V_{1}$ and $V_{2}$ are substantially less than
the characteristic scale of $\hat{H}_{0}$, we can neglect the $\mathbf{R}$-dependence
of the Berry autocorrelation function, substituting $C_{\mathrm{Berry}}(r,\mathbf{R})\approx C_{\mathrm{Berry}}(r,\mathbf{R}_{0})$,
where $\mathbf{R_{0}}$ is any point in the vicinity of $V_{1}$ and
$V_{2}$. Then Fourier transform leads to
\begin{equation}
\sum_{\mathbf{n},\mathbf{n}'\in\mathrm{MC}}\left\langle \mathbf{n}|\hat{V}_{1}|\mathbf{n}'\right\rangle \left\langle \mathbf{n}'|\hat{V}_{2}|\mathbf{n}\right\rangle \approx\int d^{D}p\tilde{V}_{1}(\mathbf{p})\tilde{V}_{2}(-\mathbf{p})\widetilde{C_{\mathrm{Berry}}^{2}}(p),\label{eq:sumnnondFour}
\end{equation}
where 
\[
\tilde{V}_{i}(\mathbf{p})=(2\pi)^{-D/2}\intop d^{D}rV_{i}(\mathbf{r})e^{-i\mathbf{pr}},\quad\widetilde{C_{\mathrm{Berry}}^{2}}(p)=(2\pi)^{D/2}\intop_{0}^{\infty}r^{D-1}dr\frac{J_{D/2-1}(pr)}{(pr)^{D/2-1}}C_{\mathrm{Berry}}^{2}(r,\mathbf{R}).
\]
In any dimension, if $V_{1}$ and $V_{2}$ can be expanded in terms
of hyperspherical harmonics, $V_{i}(\mathbf{r})$ and its Fourier
transform $\tilde{V}_{i}(\mathbf{p})$ are expanded in terms of the
same hyperspherical harmonics (see \citep{avery2012hyperspherical}).
For example, in the 3D case ($D=3$), when $V_{i}(\mathbf{r})=\sum_{l,m}v_{ilm}(r)Y_{lm}(\mathbf{r}/r)$,
its Fourier transforms can be expanded in terms of the same spherical
harmonics, $\tilde{V}_{i}(\mathbf{p})=\sum_{l,m}\tilde{v}_{ilm}(p)Y_{lm}(\mathbf{p}/p)$,
where $\tilde{v}_{ilm}(p)=\sqrt{2/\pi}(-i)^{l}\int_{0}^{\infty}r^{2}drj_{l}(pr)v_{ilm}(r)$
and $j_{l}$ is the spherical Bessel function. Substituting these
expansions into Eq. \eqref{eq:sumnnondFour} we see that, due to the
orthogonality of the spherical (or hyperspherical) harmonics, $W_{2}^{\mathrm{nond}}$
vanishes when $V_{1}$ and $V_{2}$ do not contain the same harmonics.
For example, it is the case when $V_{1}$ and $V_{2}$ are different
terms in multipole expansion of a potential. The odd (even) $V_{i}(\mathbf{r})$
includes only spherical harmonics with odd (even) $l$. Then for the
perturbations of different parity the condition (\Wapp) is satisfied
too.

The pseudopotential expansion of multipolar interactions \citep{derevianko2003,*derevianko2005}
has the form
\begin{equation}
\hat{V_{i}}=\sum_{l,m,l',m'}\left|v'_{il'm'}(r)Y_{l'm'}(\mathbf{r}/r)\left\rangle \right\langle v{}_{ilm}(r)Y_{lm}(\mathbf{r}/r)\right|.\label{eq:Vpseudo}
\end{equation}
The sum in $W_{2}^{\mathrm{nond}}$ contains the matrix elements $\left\langle v_{1lm}(r_{1})Y_{lm}(\mathbf{r}_{1}/r_{1})\left|C_{\mathrm{Berry}}(|\mathbf{r}_{1}-\mathbf{r}_{2}|,\mathbf{R}_{0})\right|v_{2l''m''}(r_{2})Y_{l''m''}(\mathbf{r}_{2}/r_{2})\right\rangle \propto\delta_{l''l}\delta_{m''m}$
{[}since the matrix elements have the same form as Eq. \eqref{eq:sumnnond},
the derivation above is applicable{]}. Then different terms in the
expansion \eqref{eq:Vpseudo} can be considered as independent perturbations.

Equation \eqref{eq:Wdiag2V} also shows that $W_{s}(E_{\mathbf{n}},E_{\alpha_{s}})$
should decay as $\left(E_{\mathbf{n}}-E_{\alpha_{s}}\right)^{-2}$
in the limit $\left|E_{\mathbf{n}}-E_{\alpha_{s}}\right|\rightarrow\infty$.

\section{Derivation of the recurrence relation (\NPCs) for NPC }

Choosing the Lorentzian profile (\Lorentz) for the continuous strength
function $W_{s}(E_{\mathbf{n}},E_{\alpha_{s}})\thickapprox W_{L}(E_{\alpha_{s}}-E_{\mathbf{n}},\Gamma_{s})\Delta E$,
replacing the sum over $E_{\alpha_{s-1}}$ by the integral, and substituting
$\left\langle \left|\left\langle \alpha_{s}|\alpha_{s-1}\right\rangle \right|^{2}\right\rangle =W'(E_{\alpha_{s}},E_{\alpha_{s-1}})\Delta E$,
we reduce (\Wssm) to the integral equation:
\begin{equation}
W_{L}(E_{\alpha_{s}}-E_{\mathbf{n}},\Gamma_{s})=C_{SI}\int dE_{\alpha_{s-1}}W'(E_{\alpha_{s}},E_{\alpha_{s-1}})W_{L}(E_{\alpha_{s-1}}-E_{\mathbf{n}},\Gamma_{s-1}),\label{eq:inteqWp}
\end{equation}
where the factor $C_{SI}$ compensates for inaccuracy of the sum-integral
replacement. The solution of \eqref{eq:inteqWp} $W'(E_{\alpha_{s}},E_{\alpha_{s-1}})=W_{L}(E_{\alpha_{s}}-E_{\alpha_{s-1}},\Gamma')/C_{SI}$
with $\Gamma'=\Gamma_{s}-\Gamma_{s-1}$ is obtained using the Fourier
transform.

Whenever it is applicable, the Fermi golden rule provides \citep{wigner1955,fyodorov1995,frahm1995,jacquod1995,Olshanii2012}
\[
\Gamma_{s}\approx\frac{\pi}{\mathcal{N}_{\mathrm{MC}}^{2}\Delta E}\sum_{\mathbf{n},\mathbf{n}'\in\mathrm{MC}}\left|\left\langle \mathbf{n}\left|\sum_{s'=1}^{s}\hat{V}_{s'}\right|\mathbf{n}'\right\rangle \right|^{2}.
\]
For independent perturbations, obeying the condition (\Wapp), the
sum over $\mathbf{n}$ and $\mathbf{n}'$ above can be split as
\[
\sum_{\mathbf{n},\mathbf{n}'\in\mathrm{MC}}\left|\left\langle \mathbf{n}\left|\sum_{s'=1}^{s}\hat{V}_{s'}\right|\mathbf{n}'\right\rangle \right|^{2}=\sum_{\mathbf{n},\mathbf{n}'\in\mathrm{MC}}\left|\left\langle \mathbf{n}\left|\sum_{s'=1}^{s-1}\hat{V}_{s'}\right|\mathbf{n}'\right\rangle \right|^{2}+\sum_{\mathbf{n},\mathbf{n}'\in\mathrm{MC}}\left|\left\langle \mathbf{n}\left|\hat{V}_{s}\right|\mathbf{n}'\right\rangle \right|^{2}.
\]
Then $\Gamma_{s}=\Gamma_{s-1}+\Gamma'$, in the agreement with the
general relation derived above.

Using Eq. \eqref{eq:FEDa}, we can represent IPR as the average over
states with a fixed energy difference
\begin{equation}
\eta_{s}\equiv\sum_{\mathbf{n}}\overline{\left|\left\langle \alpha_{s}|\mathbf{n}\right\rangle \right|^{4}}^{\alpha_{s}}=\sum_{\mathbf{n}}\left\langle \left|\left\langle \alpha_{s}|\mathbf{n}\right\rangle \right|^{4}\right\rangle .\label{eq:IPRaver}
\end{equation}
According to Berry's conjecture \citep{berry1977}, non-integrable
system eigenfunctions are Gaussian random functions. The Gaussian
distribution of the $\left\langle \mathbf{r}|\alpha_{s}\right\rangle $
values was confirmed even for single-scatterer models \citep{seba1990,Stone2010}
which have IPR $\sim0.5$ and, therefore, are rather far from the
eigenstate thermalization regime. As a consequence, $\left\langle \mathbf{n}|\alpha_{s}\right\rangle =\int d^{D}r\left\langle \mathbf{n}|\mathbf{r}\right\rangle \left\langle \mathbf{r}|\alpha_{s}\right\rangle $
can be considered as random Gaussian variables. In T-invariant systems,
$\left\langle \mathbf{r}|\alpha_{s}\right\rangle $ and $\left\langle \mathbf{r}|\mathbf{n}\right\rangle $
can always be chosen to be everywhere real. Therefore, $\left\langle \mathbf{n}|\alpha_{s}\right\rangle $
are real random Gaussian variables. In PT-invariant systems, $\left\langle \mathbf{-r}|\alpha_{s}\right\rangle ^{*}=\pm\left\langle \mathbf{r}|\alpha_{s}\right\rangle $
and $\left\langle \mathbf{-r}|\mathbf{n}\right\rangle ^{*}=\pm\left\langle \mathbf{r}|\mathbf{n}\right\rangle $
(the signs $\pm$ in two identities are independent). Then we have
either $\left\langle \mathbf{n}|\alpha_{s}\right\rangle ^{*}=\left\langle \mathbf{n}|\alpha_{s}\right\rangle $,
$\mathrm{Im}\left\langle \mathbf{n}|\alpha_{s}\right\rangle =0$,
and $\left\langle \mathbf{n}|\alpha_{s}\right\rangle $ being a real
random Gaussian variable, or $\left\langle \mathbf{n}|\alpha_{s}\right\rangle ^{*}=-\left\langle \mathbf{n}|\alpha_{s}\right\rangle $,
$\mathrm{Re}\left\langle \mathbf{n}|\alpha_{s}\right\rangle =0$,
and $\left\langle \mathbf{n}|\alpha_{s}\right\rangle $ being a real
random Gaussian variable multiplied by $i$. If the system is not
T-invariant nor PT-invariant $\left\langle \mathbf{n}|\alpha_{s}\right\rangle $
should be considered as complex random Gaussian variables.

The fourth moment of a Gaussian variable is the second one squared
and multiplied by 2 or 3 for complex or real variables, respectively
(see \citep{Olshanii2012}). Then averaged IPR \eqref{eq:IPRaver}
can be transformed as

\begin{equation}
\eta_{s}=\left\{ \begin{array}{c}
2\\
3
\end{array}\right\} \sum_{\mathbf{n}}\left\langle \left|\left\langle \alpha_{s}|\mathbf{n}\right\rangle \right|^{2}\right\rangle ^{2}\thickapprox\left\{ \begin{array}{c}
2\\
3
\end{array}\right\} \int dE_{\mathbf{n}}W_{L}^{2}(E_{\alpha_{s}}-E_{\mathbf{n}},\Gamma_{s})\Delta E=\left\{ \begin{array}{c}
2\\
3
\end{array}\right\} \frac{\Delta E}{2\pi\Gamma_{s}}\label{eq:etasGamma-1}
\end{equation}
This leads to the relation for NPC (\NPCs) with $\nu=2\pi\Gamma'/(2\Delta E)$
or $\nu=2\pi\Gamma'/(3\Delta E)$, respectively. For large $s$ the
chaotic properties of $\left|\alpha_{s}\right\rangle $ weakly depend
on $s$ and if the shape of $\hat{V}_{s}$ is independent of $s$
(as for scatterers with the same strength), $\nu$ will be approximately
independent of $s$. 

\section{harmonic waveguide with zero-range scatterers along its axis}

Projecting the Schr\"odinger equation with the rank-$s$ separable
potential
\[
\left(\hat{H}_{0}+\sum_{s'=1}^{s}V_{s'}\left|\mathcal{F}_{s'}\right\rangle \left\langle \mathcal{F}_{s'}\right|\right)\left|\alpha_{s}\right\rangle =E_{\alpha_{s}}\left|\alpha_{s}\right\rangle 
\]
onto eigenstates $\left|\mathbf{n}\right\rangle $ of $\hat{H}_{0}$,
we get the expansion coefficients (\nalpha). Then we can express
$\left|\alpha_{s}\right\rangle $ in terms of the overlaps $\left\langle \mathcal{F}_{s'}|\alpha_{s}\right\rangle $,
\begin{equation}
\left|\alpha_{s}\right\rangle =\sum_{\mathbf{n}}\sum_{s'=1}^{s}\frac{\left|\mathbf{n}\right\rangle \left\langle \mathbf{n}|\mathcal{F}_{s'}\right\rangle }{E_{\alpha}-E_{\mathbf{n}}}V_{s'}\left\langle \mathcal{F}_{s'}|\alpha_{s}\right\rangle .\label{eq:alpha}
\end{equation}
Finally, projection of this equation onto the formfactors $\left|\mathcal{F}_{s''}\right\rangle $
leads to the set of linear equations (\Fsalpha) for the overlaps.

Eigenfunctions and eigenenergies of $\hat{H}_{0}$ (\Hzero) are expressed
as
\begin{equation}
\left\langle \rho,z|nl\right\rangle =L^{-1/2}e^{2i\pi lz/L}\left\langle \rho|n\right\rangle ,\quad E_{nl}=\hbar\omega_{\perp}(2n+1)+\hbar^{2}(2\pi l/L-A)^{2}/(2m)\label{eq:zlpbc}
\end{equation}
with $-\infty<l<\infty$ for PBC and
\begin{equation}
\left\langle \rho,z|nl\right\rangle =(2/L)^{1/2}\left\langle \rho|n\right\rangle \sin\pi lz/L,\quad E_{nl}=\hbar\omega_{\perp}(2n+1)+(\pi\hbar l/L)^{2}/(2m)\label{eq:zlhwb}
\end{equation}
with $1\leq l<\infty$ for the hard-wall box. Here
\begin{equation}
\left\langle \rho|n\right\rangle =\frac{1}{\sqrt{\pi}a_{\perp}}L_{n}^{(0)}((\rho/a_{\perp})^{2})\exp(-(\rho/a_{\perp})^{2}/2)\label{eq:rhon}
\end{equation}
is the radial wavefunction, $a_{\perp}=(\hbar/m\omega_{\perp})^{1/2}$
is the transverse oscillator range, and $L_{n}^{(0)}$ are the Laguerrre
polynomials (see \citep{DLMF}). Now $\mathbf{n}=(n,l)$ and, like
in \citep{Yurovsky2010}, the sum over $l$ in (\Fsalpha) with the
eigenfunctions and eigenenergies \eqref{eq:zlpbc} can be calculated
analytically using the formula 
\begin{equation}
\sum_{l=-\infty}^{\infty}\frac{\exp(2i\pi l\zeta)}{l+a}=\frac{\pi}{\sin\pi a}\exp(-2i\pi a(\zeta-[\zeta]-1/2))\label{eq:summlexp}
\end{equation}
(see Eq. (5.4.3.4) in \citep{prudnikov}). Then the system (\Fsalpha)
attains the form
\begin{equation}
\sum_{s''=1}^{s}S_{s's''}(\varepsilon)\left\langle \mathcal{F}_{s''}|\alpha_{s}\right\rangle =0\label{eq:linsysreg}
\end{equation}
with $S_{s's''}(\varepsilon)=\left\langle \mathcal{F}_{s'}|S(z-z_{s''},\varepsilon)\right\rangle V_{s''}/V_{0}-\delta_{s's''}$.
Here $S(z,\varepsilon)$ is the Green function of the integrable system
on the waveguide axis. For $z\geq0$ it is expressed as
\begin{equation}
S(z,\varepsilon)=\sqrt{\lambda}\sum_{n=0}^{\infty}\frac{1}{2p_{n}}e^{2i\pi l_{0}z/L}\left[e^{2ip_{n}z/L}\left(\cot(\pi l_{0}+p_{n})-i\right)-e^{-2ip_{n}z/L}\left(\cot(\pi l_{0}-p_{n})-i\right)\right],\label{eq:Sze}
\end{equation}
where $\varepsilon=mL^{2}(E-\hbar\omega_{\perp})/(2\hbar^{2})$ and
$p_{n}=\sqrt{\varepsilon-\lambda n}$. For $z<0$ we have $S(-z,\varepsilon)=S^{*}(z,\varepsilon)$.
The formfactors $\left\langle \mathbf{r}|\mathcal{F}_{s'}\right\rangle =\delta_{\mathrm{reg}}(\mathbf{r}-\mathbf{R}_{s'})$
are expressed in terms of the Fermi-Huang pseudopotential $\delta_{\mathrm{reg}}(\mathbf{r})f(\mathbf{r})=\frac{\partial}{\partial r}\left[rf(\mathbf{r})\right]_{\mathbf{r}=0}$
which extracts the regular part of the function $f(\mathbf{r})$ at
$r\rightarrow0$. Then for $s'\neq s''$ we have $S_{s's''}(\varepsilon)=S(z_{s'}-z_{s''},\varepsilon)$
since $S(z_{s'}-z_{s''},\varepsilon)$ is regular in this case. However,
$S(z,\varepsilon)$ has a pole at $z\rightarrow0$. The asymptotic
of the irregular part at $z\rightarrow0$
\[
-\sqrt{\lambda}\sum_{n=[\varepsilon/\lambda]+1}^{\infty}\frac{e^{-2|p_{n}|z/L}}{|p_{n}|}\sim-\frac{L}{\sqrt{\lambda}z}-\zeta\left(\frac{1}{2},\left[\frac{\varepsilon}{\lambda}\right]+1-\frac{\varepsilon}{\lambda}\right)
\]
is expressed, as in \citep{moore2004} in terms of the Hurwitz zeta
function (see \citep{DLMF}). The term $\sim1/z$ is eliminated by
the Fermi-Huang pseudopotential. Then we can represent the diagonal
matrix elements as
\[
S_{s's'}(\varepsilon)=\frac{V_{s'}}{V_{0}}\left[\sqrt{\lambda}\sum_{n=0}^{\infty}\left(\frac{\sin2p_{n}}{p_{n}(\cos2\pi l_{0}-\cos2p_{n})}+\frac{\theta(n-\varepsilon/\lambda)}{|p_{n}|}\right)-\zeta\left(\frac{1}{2},\left[\frac{\varepsilon}{\lambda}\right]+1-\frac{\varepsilon}{\lambda}\right)\right]-1.
\]

In the T-invariant model, $A=l_{0}=0$ and the function \eqref{eq:Sze}
is real. 

The case of a hard-wall box is treated in a similar way. For $s'\geq s''$
we have
\[
S_{s's''}(\varepsilon)=-\frac{V_{s''}}{V_{0}}\left[\sqrt{\lambda}\sum_{n=0}^{\infty}\left(2\frac{\sin2p_{n}(1-z_{s'}/L)\sin2p_{n}z_{s''}/L}{p_{n}\sin2p_{n}}-\frac{\theta(n-\varepsilon/\lambda)}{|p_{n}|}\delta_{s's''}\right)+\zeta\left(\frac{1}{2},\left[\frac{\varepsilon}{\lambda}\right]+1-\frac{\varepsilon}{\lambda}\right)\delta_{s's''}\right]-\delta_{s's''}
\]
and $S_{s''s'}(\varepsilon)=S_{s's''}(\varepsilon)$.

The homogeneous system of linear equations \eqref{eq:linsysreg} has
non-trivial solutions if the determinant of its matrix is equal to
zero. Then the roots $\varepsilon_{\alpha}$ of the determinant, $\det(S_{s's''}(\varepsilon_{\alpha}))=0$,
give us the non-integrable system eigenenergies $E_{\alpha}=2\hbar^{2}\varepsilon_{\alpha}/(mL^{2})+\hbar\omega_{\perp}$
and solutions of the system \eqref{eq:linsysreg}, being substituted
to Eq. \eqref{eq:alpha}, provide the eigenfunctions. 

\end{widetext}
\end{document}